\documentclass[fleqn,usenatbib]{mnras}
\usepackage{newtxtext,newtxmath}
\usepackage{arydshln}
\usepackage[T1]{fontenc}
\DeclareRobustCommand{\VAN}[3]{#2}
\let\VANthebibliography\thebibliography
\def\thebibliography{\DeclareRobustCommand{\VAN}[3]{##3}\VANthebibliography}
\usepackage{anyfontsize}

\usepackage{graphicx}	
\usepackage{amsmath}	
\usepackage{xcolor}
\usepackage{ulem}
\newcommand{\pc}{\,{\rm pc}\,}
\newcommand{\kpc}{\,{\rm kpc}\,}
\newcommand{\Myr}{\,{\rm Myr}\,}
\newcommand{\Gyr}{\,{\rm Gyr}\,}
\newcommand{\kms}{\,{\rm km s}$^{-1}$\,}
\newcommand{\G}{\,{\rm G}\,}
\newcommand{\muG}{\,$\mu$\G\,}
\newcommand{\erg}{\,{\rm erg}\,}

\newcommand{\BB}{\mathbf{B}}
\newcommand{\JJ}{\mathbf{J}}
\newcommand{\bb}{\mathbf{b}}
\newcommand{\UU}{\mathbf{U}}
\newcommand{\uu}{\mathbf{u}}

\newcommand{\emf}{\mbox{\boldmath ${\cal E}$} {}}
\newcommand{\mean}[1]{\overline{#1}}

\newcommand{\eref}{Eq.~\,\ref}
\newcommand{\fref}{Fig.~\,\ref}
\newcommand{\sref}{Section~\,\ref}
\newcommand{\aref}{Appendix~\,\ref}

\newcommand{\comment}[1]{}
\long\def\/*#1*/{}

\title[IROS method]{Iterative removal of sources to model the turbulent electromotive force}

\author[A. B. Bendre et al.]{
Abhĳit B. Bendre$^{1,5}$\thanks{E-mail: abhijit.bendre@sns.it},
Jennifer Schober$^{1}$,
Prasun Dhang$^{2,4}$ and 
Kandaswamy Subramanian$^{2,3}$
\\
$^{1}$Institute of Physics, Laboratory of Astrophysics, \'Ecole Polytechnique F\'ed\'erale de Lausanne (EPFL), 1290 Sauverny, Switzerland\\
$^{2}$IUCAA, Post Bag 4, Ganeshkhind, Pune 411007, India \\
$^{3}$Department of Physics, Ashoka University, Rajiv Gandhi Education City, Rai, Sonipat 131029, Haryana, INDIA\\
$^{4}$JILA, University of Colorado and National Institute of Standards and Technology, 440 UCB, Boulder, CO 80309-0440, USA\\
$^{5}$Scuola Normale Superiore di Pisa, Piazza dei Cavalieri 7, 56126 Pisa, Italy
}

\date{Accepted XXX. Received YYY; in original form ZZZ}

\pubyear{2015}
\begin{document}
\label{firstpage}
\pagerange{\pageref{firstpage}--\pageref{lastpage}}
\maketitle

\begin{abstract}
We describe a novel method to compute the components of dynamo tensors 
from direct magnetohydrodynamic (MHD) simulations. Our method relies 
upon an extension and generalisation of the standard H\"ogbom CLEAN 
algorithm widely used in radio astronomy to systematically remove the 
impact of the strongest beams onto the corresponding image. This 
generalisation, called the Iterative Removal of Sources (IROS) method, 
has been adopted here to model the turbulent electromotive force (EMF) 
in terms of the mean magnetic fields and currents. Analogous to the 
CLEAN algorithm, IROS treats the time series of the mean magnetic 
field and current as beams that convolve with the dynamo coefficients 
which are treated as (clean) images to produce the EMF time series (
the dirty image). We apply this method to MHD simulations of galactic 
dynamos, to which we have previously employed other methods of computing 
dynamo coefficients such as the test-field method, the regression 
method, as well as local and non-local versions of the singular value 
decomposition (SVD) method. We show that our new method reliably 
recovers the dynamo coefficients from the MHD simulations. It also 
allows priors on the dynamo coefficients to be incorporated easily 
during the inversion, unlike in earlier methods. Moreover, using 
synthetic data, we demonstrate that it may serve as a viable 
post-processing tool in determining the dynamo coefficients, even when 
the power of additive noise to the EMF is twice as much the actual EMF.
\end{abstract}




\section{Introduction}
\label{sec:intro}
Magnetic fields with long-range regularity are present in various 
astrophysical systems from stars to galaxies. Observational and 
numerical evidence suggests that the most plausible mechanism for 
their generation is the large-scale dynamo mechanism \citep{mof_book,
beck_wielebinski,Rudiger,BS05,anv_kan_book}. In this process, the 
magnetic field with scales of regularity much larger than the scale 
at which turbulence is driven, is generated at the expense of kinetic 
energy, with the aid of both large-scale shear and helical turbulence. 
This mechanism can be characterised in terms of mean-field 
electrodynamics \citep{kr_rad_book}, by first expressing the dynamical 
variables, velocity $\UU$ and magnetic field $\BB$, as the sums of their 
respective mean or large-scale components ($\mean{\UU}$ and $
\mean{\BB}$) and the fluctuating (or the small-scale) components ($\uu$
and $\bb$).  The mean is defined as an average over a suitable domain for 
which the Reynolds's averaging rules are satisfied. Which is to say that; $
\mean{\partial \BB/\partial t}= \partial \mean{\BB}/\partial t,\,
\mean{\partial \BB/\partial x_i}= \partial\mean{ \BB}/\partial x_i,
\, \mean{\BB_1+\BB_2} = \mean{\BB}_1 +\mean{\BB}_2,\,\mean{\mean{
\BB}}=\mean{\BB},\, \mean{\mean{B}_{i}b_j}=0,\, \mean{\mean{B}_i
\mean{B}_{j}}=\mean{B}_{i}\mean{B}_{j}$.  The evolution of the mean 
magnetic field then depends on the mean velocity and turbulent EMF, i.e. the 
cross correlation between $\uu$ and $\bb$, $\mean{\emf}_i=(\mean{\uu
\times\bb})_i$. This EMF is then expressed in terms of the mean magnetic 
field itself and the system is closed, i.e. 
\begin{align}
\mean{\emf}_i=\alpha_{ij}
\mean{B}_j-\eta_{ij}\,(\nabla\times\mean{\BB})_j,
\label{eq:emf_local}
\end{align}
when mean fields are defined using horizontal averaging\footnote{We 
also note that we have used a convention such that the repeated indices are
summed over.} (see \eref{eq:mean}). Here $\alpha_{ij}$ and $\eta_{ij}
$ are the coefficients of the dynamo tensors, which in general relate to the
statistical properties of the background turbulence. In the high conductivity
limit, these coefficients are proportional to the specific turbulence 
properties such as the fluctuating kinetic and magnetic energy densities and 
their helicity densities. Therefore in the context of various astrophysical 
systems, these dynamo coefficients dictate how statistical properties of the 
background turbulence impact the evolution of large-scale magnetic fields. 
Observational estimates of the dynamo coefficients often rely upon 
assumptions and phenomenological properties of the turbulence. In the 
context of galactic dynamo, e.g., the estimation relies upon the observables 
such as correlation length scales of turbulence and the supernova (SN) rate 
\citep{kr_rad_book,anv_kan_book}. Equally, analytical estimates of the
transport coefficients \citep[e.g.][]{CH96} make simplifying assumptions (such as
isotropy of turbulence and high conductivity etc.) that are often 
questionable. Therefore direct magnetohydrodynamic (MHD) simulations of
such systems with realistic turbulence driving and wherein the $\mean{
\emf}$ and $\mean{\BB}$ are self-consistently generated, provide a very 
useful tool in estimating the dynamo coefficients.

Various methods to compute these dynamo coefficients from direct 
numerical simulations (DNS) have so far been suggested and tested. The 
plethora of methods reflects the fact that \eref{eq:emf_local} represents
an under-determined system and one that is not straightforward to invert. 
For example, \cite{CH96} used the random magnetic field generated in helically 
driven turbulence with uniform imposed mean field to calculate the EMF 
and fit it against the imposed $\mean{\mathbf{B}}$ to compute the $
\alpha_{ij}$ coefficients, while in \cite{angstrom} the authors used a method
of determining the conductivity of solids from material science to determine 
the magnetic diffusivity. In a slightly different setup \cite{Park_2023} explored 
the scenario of magnetically forced turbulence (with explicit helical 
magnetic forcing), and computed turbulent diffusivity using a 
semi-analytical approach. This is relevant for magnetic field amplification in
magnetically dominated systems such as accretion disks, stellar coronas etc.
In \cite{BranSok02} and \cite{Kowal06}, the authors developed an approach to 
computing the dynamo coefficients by fitting the various moments of the 
EMF and mean fields against their respective linear relations. This method is 
designed to handle the additive noise in the EMF and mean magnetic field 
data assuming that they are uncorrelated. A method with similar capability 
and an additional advantage of quantifying the auto-correlations between 
the dynamo coefficients, uses singular value decomposition (SVD), and was 
applied to several contexts. These include the stellar dynamo simulations in
\cite{racine2011mode} and \cite{simard2016characterisation}, the interstellar
medium (ISM) simulations in \cite{loc_svd}, and the global accretion disc
simulations in \cite{prasun_2020}. The same method was extended to explore
the non-local dependence of EMF on mean magnetic fields in 
\cite{non_loc_svd} and to compute the scale-dependent dynamo coefficients in
the ISM simulations.

As a more direct approach to invert \eref{eq:emf_local}, the kinematic 
test-field method was introduced by \citep{schriner_test,schriner_test1}. 
In this method the additional test magnetic fields $\mean{\BB}_T$ are 
passively evolved along with the DNS such that they do not affect the 
turbulent velocity $\mean{\uu}$. The fluctuations ($\bb_T$) generated 
through tangling of the test-fields by the turbulence, are recorded to further
compute the additional EMF components, and so also the dynamo
coefficients. The test-field method has an advantage of determining the 
dynamo coefficients at the length scale of imposed the test-fields, and it has
been used in several different contexts, including helically driven
turbulence, ISM turbulence, accretion disc turbulence, and even in the 
simulations of solar and geo-dynamos \citep{Bran05,sur2007kinetic,
gressel_2008,kapala_test,bendre2015dynamo,GP15,War17}. However,
it often proves to be computationally expensive as it relies upon integrating
the additional induction equations (associated with each test-field
component) along with the rest of the MHD equations. Pros and cons of
these methods have also been discussed in the literature
\citep{Brandenburg2009,brandenburg_2018,anv_kan_book}.

In this paper we introduce a novel method to determine the dynamo 
coefficients which is based on post-processing the data from turbulent
dynamo DNS. For this we adopt a version of a deconvolution algorithm 
based on the H\"ogbom CLEAN method \citep{hogbom74} called IROS (Iterative 
Removal Of Sources) \citep{iros} to mitigate some of the issues in previous 
methods. This method is then tested against mock data and also applied to 
determine dynamo coefficients from a DNS of ISM simulations, for which 
comparison can be made to the results from several of the other methods. 
We have also successfully applied this method to compute the dynamo 
coefficients for other systems as well \citep[eg.][]{prasun,yasin}, where the 
turbulence itself is driven by magnetic instabilities. In
\sref{sec:mock_data}, we apply the aforementioned methods to the mock 
data of EMF with an additive random noise. Further in \sref{sec:results} 
we apply these methods to the data of ISM simulations, which is followed by
\sref{sec:conclusion}, where we discuss the conclusions.

This paper is structured as follows. In \sref{sec:simulations} we describe
briefly the details of the ISM simulations, and the data which we use in the
later sections to apply the IROS method. In \sref{sec:iros_method} we
describe the local version of the IROS method. It if followed by
\sref{sec:cIROS}, where we explore the possibility of imposing physical 
constraints on the dynamo coefficients in the framework of IROS. In this 
analysis, we impose the constraints of symmetry or antisymmetry on the 
vertical profiles of these dynamo coefficients. Further in \sref{sec:results}
we apply these methods to the data of ISM simulations, which is followed by
\sref{sec:conclusion}, where we discuss the conclusions. The paper also 
has two appendices in which we describe respectively the application of 
IROS method to invert the non-local representation of EMF 
(\aref{sec:nonlocal}),  and the one in which we represent the IROS method 
as a flowchart (\aref{sec:flowchart}).

\begin{figure}
    \centering
    \includegraphics[width=\linewidth]{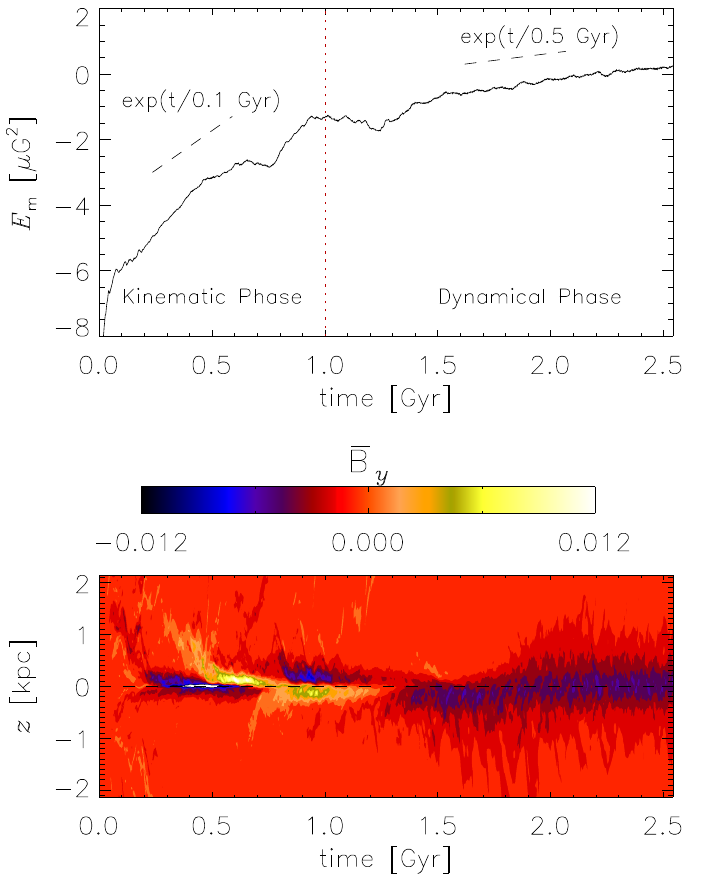}
    \caption{In the top panel, time evolution of the mean magnetic energy 
            is shown and kinematic and the dynamical phases of magnetic 
            field evolution are separated by a vertical red dotted line. 
            In the bottom the panel time evolution of the vertical profile 
            of mean magnetic field ($y$-component) is shown. Colour-code 
            is normalised with a factor of $\exp{(-t/0.2\,{\rm Gyr})}$ in 
            the kinematic phase and with $\exp{(-t/1.0\,{\rm Gyr})}$ in 
            the dynamical phase to compensate for the exponential 
            amplification.}
    \label{fig:em_vs_t_and_contour}
\end{figure}
\section{Simulations}
\label{sec:simulations}
As a test case for our new method of determining dynamo coefficients 
from DNS, we use the data from the direct MHD simulations of a 
fraction of a disc galaxy. Details of these simulation setups and 
their outcomes are described in \cite{bendre2015dynamo} and 
\cite{bendre_thesis}. We describe them here briefly, for 
completeness. These were local shearing Cartesian box simulations 
of a vertically stratified distribution of ISM; performed with the 
\texttt{NIRVANA} MHD code \citep{nirvana}. The computational 
domain represented a patch of the galactic disk with properties 
consistent with those in the solar neighbourhood. Radial ($x$) and 
azimuthal ($y$) extent of this box was roughly 1\kpc by 1\kpc while 
it extended from $\sim-2$ to 2 \kpc on either sides of galactic 
midplane. Shearing periodic and periodic boundary conditions were 
used in $x$ and $y$ directions, respectively, to facilitate the 
shear and approximate axisymmetry. While the outflow conditions 
were used at the upper and lower boundaries to allow for the gas 
outflows, but to restrict the gas inflow. A flat rotation curve 
was implemented by having a radially dependent angular velocity 
$\Omega\propto R^{-1}$, such that $\Omega_0$ at the centre of 
the domain was 100\kms\kpc$^{-1}$. Turbulence was driven by the 
SN explosions modelled as the local injections of thermal energy 
($\sim 10^{51}$\erg per explosion) at a predefined rate of $\sim
7.5$\kpc$^{-2}$\Myr$^{-1}$.

With this setup the mean \footnote{Here the mean is defined as an 
average over the $x-y$ plane as defined in \eref{eq:mean}, and it 
also follows the Reynold's rules mentioned in \sref{sec:intro}}
magnetic field energy exponentially amplified with an e-folding 
time of $\sim100$\Myr for a \Gyr, until it reached equipartition 
with the turbulent kinetic energy. This growth drastically slowed 
down afterwards. We termed the initial exponential amplification 
phase (up to $\sim1$\Gyr) as the kinematic phase and the latter 
as the dynamical phase of magnetic field evolution, wherein the 
mean magnetic field is dynamically significant to affect the 
turbulent motions. Vertical profiles of the components of mean 
magnetic field went through several sign reversals and parity 
changes throughout the kinematic phase of evolution and achieved 
a stable mode, symmetric about the galactic mid-plane, with a 
strength of $\sim$\muG in the mid-plane. In 
\fref{fig:em_vs_t_and_contour} the time evolution of mean magnetic 
energy is shown along with the evolution of the vertical profile 
of $y$ component of mean magnetic field. In \citet{bendre2015dynamo} 
we have demonstrated that the initial exponential amplification 
phase of the magnetic field can be described as a solution of an 
$\alpha-\Omega$ dynamo, and that the profiles of the dynamo 
coefficients (as measured using the test-field method)  quench during 
the dynamical phase. Our estimates of the same system using the SVD
method \citep{loc_svd} also agreed largely with this assertion. Therefore to
have a reasonable comparison we use the data from the same model to 
test this new method of computing the dynamo coefficients. 

\section{Methods For Obtaining Dynamo Coefficients}
\subsection{The IROS method}
\label{sec:iros_method}
We describe here the IROS algorithm used to extract the dynamo 
coefficients associated with aforementioned MHD simulations. In 
this section, we apply it to the local representation of the EMF 
as given in \eref{eq:emf_local}. \aref{sec:nonlocal} generalises 
it further and applies the algorithm to the non-local representation. 
For this analysis we first define the mean components by averaging 
them over the $x-y$ plane, which leaves only $z$ as an independent 
variable, 
\begin{align}
\mean{\mathbf{\textbf{F}}}\left(z,t\right) =\frac{1}{L_x\,L_y} \iint \mathbf{F}\left(x,y,z,t\right)\, \mathrm{d}x\,\mathrm{d}y.
\label{eq:mean}
\end{align}
Here $\mathbf{F}$ represents any dynamical variable $\BB$, $\UU$, 
$\JJ$ ($\JJ =\nabla \times\BB$) or $\mean{\emf}$, while $L_x$ and 
$L_y$ are the sizes of the numerical domain in radial and azimuthal 
directions, respectively.

We then represent \eref{eq:emf_local} as an over-determined system 
of equations, by taking advantage of the fact that the dynamo 
coefficients do not change appreciably throughout the kinematic 
phase, since the mean field is not strong enough to affect the 
turbulence. Therefore, at any fiducial position $z=z'$ we can 
write
\begin{align}
    \mathbf{y} \left(z',t\right)= \mathbf{A}\left(z',t\right) \mathbf{x}\left(z'\right),
    \label{eq:emf_overdetermined_local}
\end{align}
such that the matrices $\mathbf{y}$ and $\mathbf{A}$ are of 
dimensions $N_t\times2$ and $N_t\times4$ respectively and
comprised of time series ($t_1\,{\rm to}\, t_N$) of both 
components of the EMF, mean magnetic fields and mean currents, 
and $\mathbf{x}$ contains the dynamo coefficients. These 
matrices are 
\begin{align}
    \mathbf{y} \left(z',t\right)&= 
    \begin{bmatrix}
    \mean{\emf}_x\left(z',t_1\right)&\mean{\emf}_y\left(z',t_1\right)\\
    \mean{\emf}_x\left(z',t_2\right)&\mean{\emf}_y\left(z',t_1\right)\\
    \vdots\\
    \mean{\emf}_x\left(z',t_N\right)&\mean{\emf}_y\left(z',t_1\right)
    \end{bmatrix},\\\nonumber\\
    \mathbf{A}^{\intercal} \left(z',t\right)&= 
    \begin{bmatrix}
     \mean{B}_x\left(z',t_1\right)  & \mean{B}_x\left(z',t_2\right) &\hdots     & \mean{B}_x\left(z',t_N\right)\\
     \mean{B}_y\left(z',t_1\right)  & \mean{B}_y\left(z',t_2\right) &\hdots     & \mean{B}_y\left(z',t_N\right)\\
    -\mean{J}_x\left(z',t_1\right)  &-\mean{J}_x\left(z',t_2\right) &\hdots     &-\mean{J}_x\left(z',t_N\right)\\
    -\mean{J}_y\left(z',t_1\right)  &-\mean{J}_y\left(z',t_2\right) &\hdots     &-\mean{J}_y\left(z',t_N\right)
    \end{bmatrix},
\end{align}
and
\begin{align}
    \mathbf{x} \left(z'\right)&= 
    \begin{bmatrix}
    \alpha_{xx}\left(z'\right)&\alpha_{yx}\left(z'\right)\\
    \alpha_{xy}\left(z'\right)&\alpha_{yx}\left(z'\right)\\
    \eta_{xx}\left(z'\right)&\eta_{yx}\left(z'\right)\\
    \eta_{xy}\left(z'\right)&\eta_{yx}\left(z'\right)
    \end{bmatrix}.
\end{align}
Unlike \eref{eq:emf_local}, this system can be solved for $\mathbf{x}
(z')$ by the least-square minimisation using the IROS method described 
below. 

\begin{figure*}
    \centering
    \includegraphics[width=0.45\linewidth]{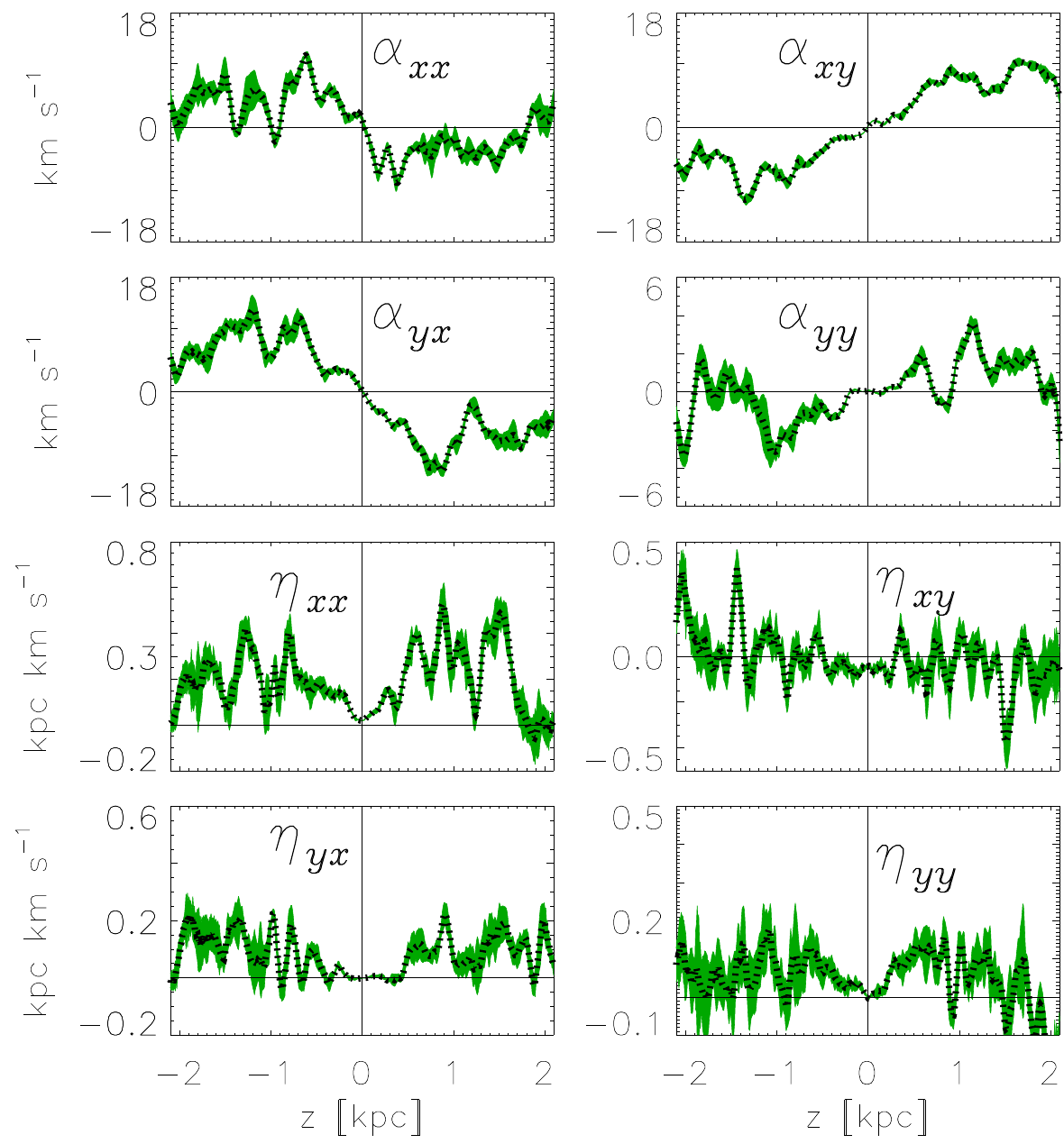}\hspace{2em}
    \includegraphics[width=0.45\linewidth]{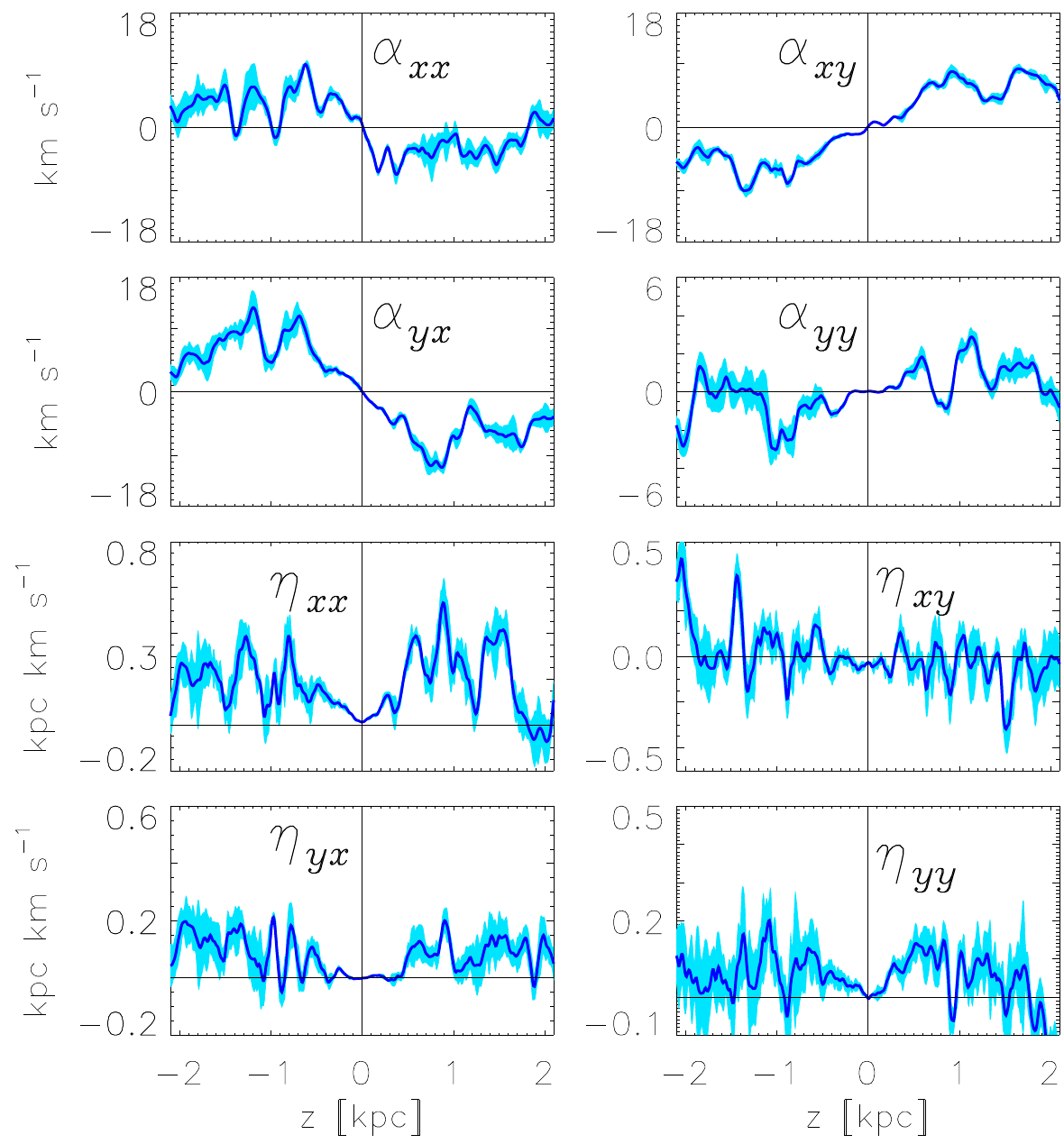}
    \caption{In left hand panels with black dotted line we 
        show the vertical profiles of dynamo coefficients 
        obtained by applying local SVD method to the data 
        of MHD simulations, along with errors which are 
        shown with green shaded region. The panels on right 
        hand side indicate the same but the local IROS method 
        has been used, the coefficients are shown with dark 
        blue solid line and the respective errors are shown 
        with regions shaded in light blue.}
    \label{fig:cofficients_local}
\end{figure*}

\subsubsection{Step 1} 
\label{sec:step1}
The IROS scheme we discuss here relies upon the incremental refinements 
to the estimates of dynamo coefficients. For that we first set all dynamo 
coefficients $\alpha_{ij}(z)$ and $\eta_{ij}(z)$ to zero. Then at any 
particular $z=z'$, we fit the time series of the $i^{\rm th}$ component 
of the EMF obtained from the simulations, $\mean{\emf}_i(z',t)$ with 
those of $\mean{B}_x(z',t)$, $\mean{B}_y(z',t)$, $\mean{J}_x(z',t)$, and 
$\mean{J}_y(z',t)$, separately (i.e. by keeping only one component of 
$\mathbf{x}_i$ non-zero and setting other components to zero) and obtain 
the zeroth level estimates $^0 \alpha_{ ix }(z') $, $^0\alpha_{iy(z')}$, 
$^0\eta_{ix}(z')$, and $\eta_{iy}(z')$ of dynamo coefficients along with 
their respective chi-squared errors ($\chi_{il}^{2}(z')$ with $l=0,1,2,4
$). Note the superscript `0' is used to indicate the zeroth level of 
refinement to the EMF. The chi-squared errors are defined as;
\begin{align}
    \chi^{2}_{i0}(z') &= \sum_t \left(\mean{\emf}_i(z',t) -\,\,^0\alpha_{ix}(z')\,\mean{B}_x(z',t)\right)^2,\nonumber\\
    \chi^{2}_{i1}(z') &= \sum_t \left(\mean{\emf}_i(z',t) -\,\,^0\alpha_{iy}(z')\,\mean{B}_y(z',t)\right)^2,\nonumber\\
    \chi^{2}_{i2}(z') &= \sum_t \left(\mean{\emf}_i(z',t) +\,\,^0\eta_{ix}(z')\,\mean{J}_x(z',t)\right)^2,\nonumber\\
    \chi^{2}_{i3}(z') &= \sum_t \left(\mean{\emf}_i(z',t) +\,\,^0\eta_{iy}(z')\,\mean{J}_y(z',t)\right)^2.
\label{eq:chi_sq_iros_old}    
\end{align}

\subsubsection{Step 2} 
\label{sec:step2}
We then subtract from the EMF of the previous step $\mean{\emf}_i
(z',t)$, the contribution corresponding to best fitted dynamo 
coefficient (the one with the least of chi-square error) 
multiplied a suitable scale-factor $\epsilon<1$. The factor 
$\epsilon$ is referred to as the ``loop gain'' in radio astronomy. 
For example, if the chi-square error associated with $^0\alpha_{iy}
(z')$ in {\it Step 1} (i.e. $\chi^{2}_{i1}(z')$) happens to be the 
smallest amongst the four, we subtract the contribution $\epsilon\,
\, ^0\alpha_{iy}(z')\,\mean{B}_y(z',t)$ from $\mean{\emf}_i(z',t)$ 
(i.e. from the EMF of the previous step) and obtain the next level 
of refinement to the fit of EMF, $^1\mean{\emf}_i(z',t)$.

\subsubsection{Step 3}
\label{sec:step3}
Only the best fitted zeroth order estimates are retained, 
while the rest are set zero. For example if $\chi^{2}_{i1}(z')$
from {\it Step 2} is the smallest, only $^0\alpha_{iy}(z')$ is 
retained, while $^0\alpha_{ix}(z')$, $^0\eta_{ix}(z')$ and $^0
\eta_{iy}(z')$ are set to zero. All dynamo coefficients 
are then updated by adding to them their zeroth level estimates 
multiplied by the loop gain.

\subsubsection{Step 4}
\label{sec:step4}
{\it Step 1} is then repeated with $^1\mean{\emf}_i(z',t)$ as the 
EMF to obtain $^1\alpha_{ix}(z')$, $^1\alpha_{iy}(z')$, $^1\eta_{
ix}(z')$, and $^1\eta_{iy}(z')$, i.e.\ the first level estimates 
of dynamo coefficients (along with their respective chi-square 
errors). The estimates of dynamo coefficients are further refined 
by adding to them the corresponding first level contribution, which 
had the least chi-square error, multiplied by the loop gain 
$\epsilon$. For example, if the chi-square associated with 
$^1\alpha_{ix}(z')$ happens to be the least, a factor of $\epsilon
\,^1\alpha_{ix}(z')$ is added to $\alpha_{ix}(z')$. The corresponding 
contribution to the EMF is again subtracted after weighting by the 
loop gain to determine the residual $^2\mean{\emf}_i$.

\begin{figure*}
    \centering
    \includegraphics[width=0.48\linewidth]{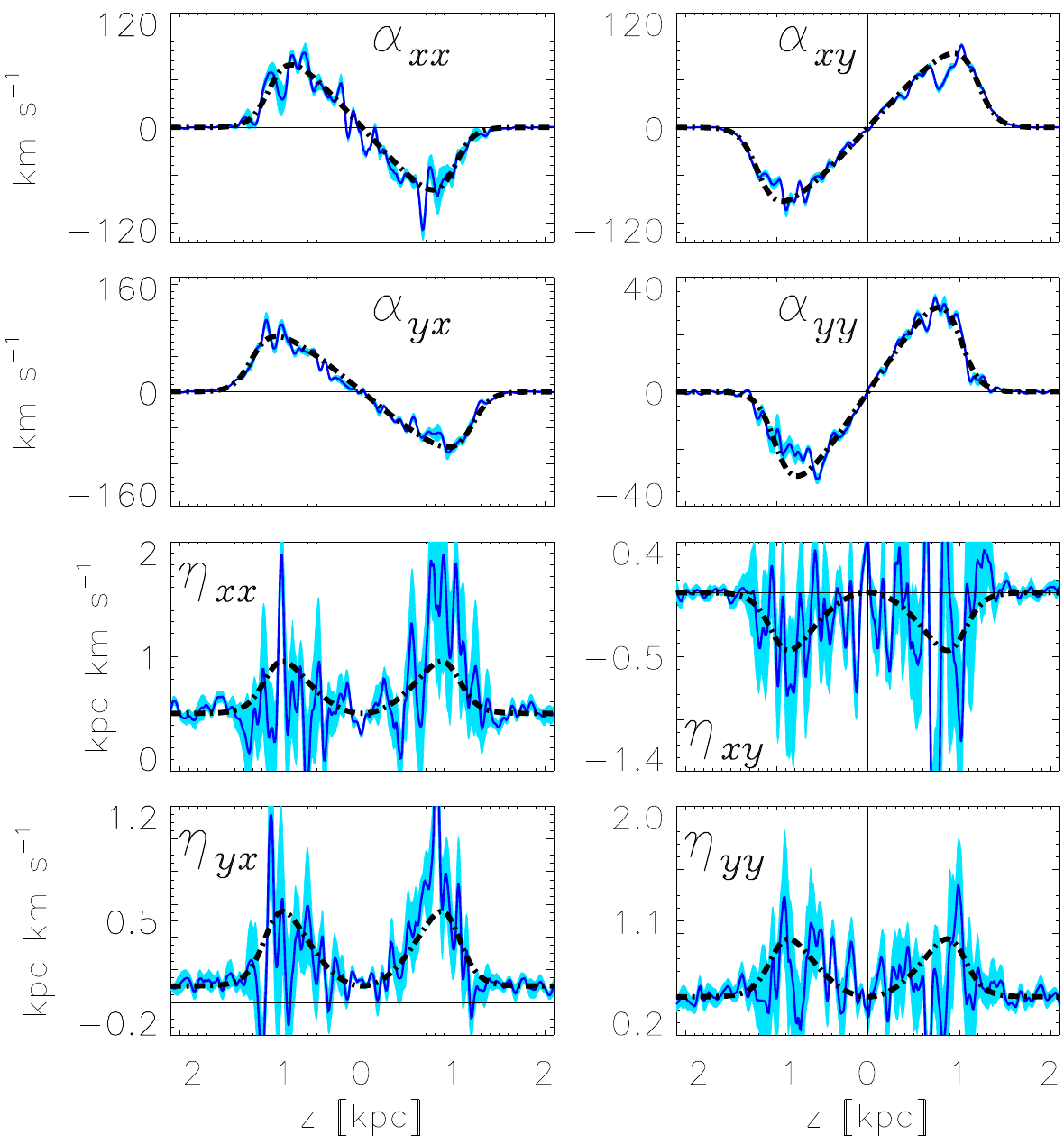}\hspace{0em}
    \includegraphics[width=0.48\linewidth]{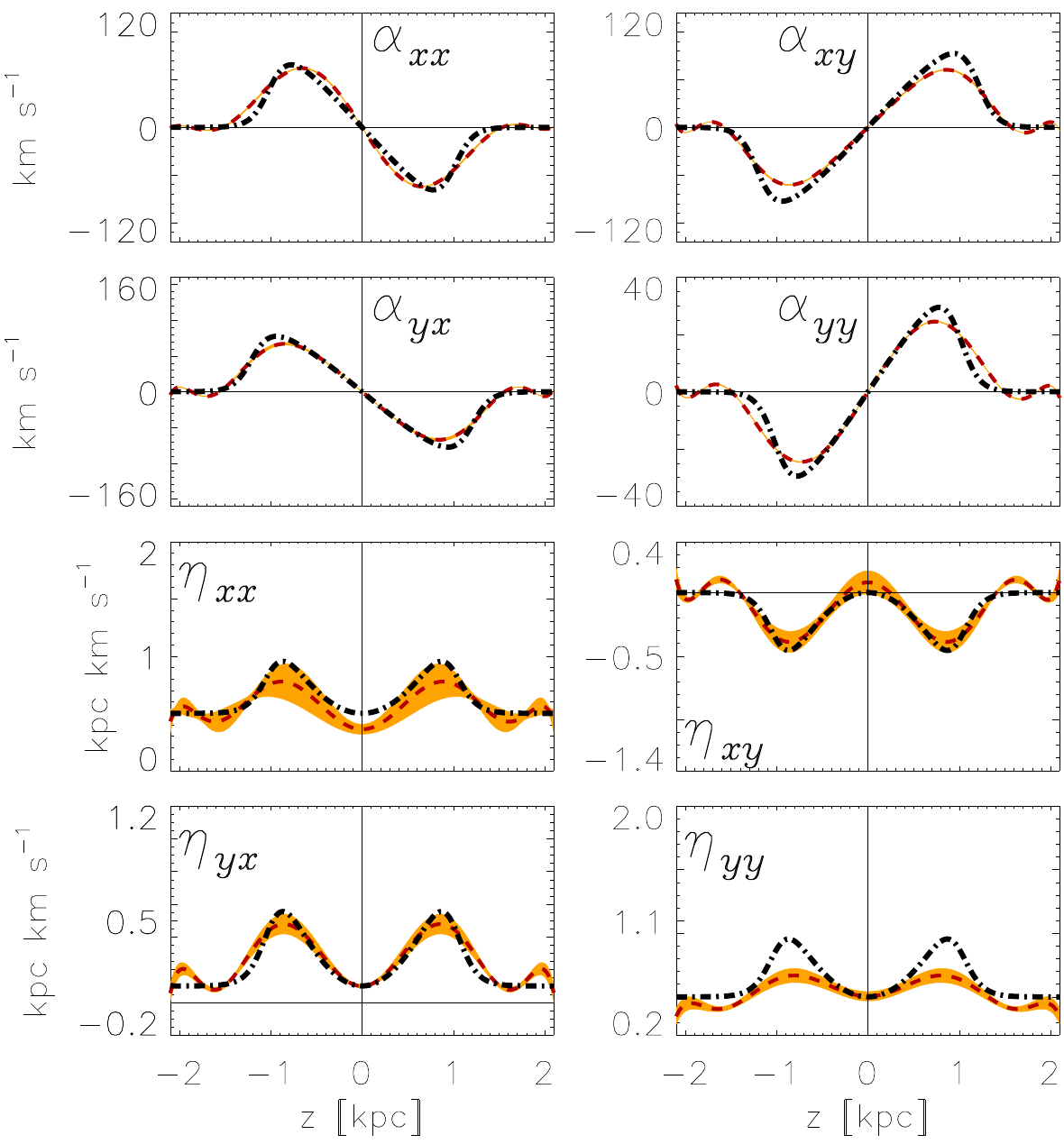}
    \caption{In the panels on the left hand side, with dark 
        blue solid lines the vertical profiles of dynamo 
        coefficients, computed from mock EMF data (with 
        additive noise) using the local IROS method are 
        shown. With light blue coloured region we have 
        indicated the $1-\sigma$ error intervals. The 
        profiles shown with black dot-dashed lines indicate 
        the dynamo coefficients used to generate the the 
        mock data. In the panels on the right hand side 
        panels the same is shown but the red dashed lines 
        indicate the coefficient recovered using cIROS 
        method, and the orange colour is used to indicate 
        the corresponding $1-\sigma$ error.}
    \label{fig:mock_data}
\end{figure*}

\subsubsection{Step 5} 
\label{sec:step5}
{\it Step 1}, {\it Step 2} and {\it Step 3} are then repeated a 
suitable number of times (say {\it R}), until the values of the 
coefficients converge or all chi-squared error values drop below 
a certain threshold (say {\it T}). Note that the coefficient 
associated with the least chi-square could be different for 
different iterations. Effectively, the final estimates for dynamo 
coefficients become; 
\begin{align}
    \alpha_{ij}(z') = \epsilon\,\,\sum_{r=0}^{R}\, ^r\alpha_{ij}(z')\\
    \eta_{ij}(z') = \epsilon\,\,\sum_{r=0}^{R}\, ^r\eta_{ij}(z') 
\end{align}
The aforementioned steps are then repeated at all $z=z'$, to 
construct the vertical profiles of all dynamo coefficients. A 
further clarification of the IROS algorithm is provided in 
\aref{sec:flowchart} using a flowchart representation of 
aforementioned steps. We describe the vertical profiles obtained 
(for MHD simulations described in \sref{sec:simulations}) using 
this algorithm in \sref{sec:results} below. Before that we also 
consider how prior constraints can be incorporated in the IROS 
reconstruction of dynamo coefficients.

\subsection{Imposing prior constraints on the dynamo coefficients}
\label{sec:cIROS}
The problem of inverting the closure for turbulent EMF can be 
equivalently treated in the framework of Bayesian inference by 
subjecting the likelihoods of the dynamo coefficients to the 
relevant priors. Such an approach is justified due to the 
helical nature of dynamo generated mean fields, which renders 
the components of mean fields and currents proportional to 
each other and thus introduces cross-correlations between 
dynamo coefficients \citep{bendre2015dynamo}. In order to then 
be able to infer the $\alpha_{ij}$ and $\eta_{ij}$ separately 
it is useful to have reasonable priors for them. Indeed, such 
physically informed priors on dynamo coefficients can be self 
consistently incorporated in the framework of IROS by putting
constraints on the profiles of the coefficients, without having 
to invoke the Bayesian approach. This is due to the fact that 
the criterion for judging the best fitted parameter to the EMF 
at each level of refinement (for \fref{fig:cofficients_local} 
and \fref{fig:coefficients_non_local} it is the least of 
chi-squared errors of the individual fits) can be chosen so 
as to impose the said constraint. For example in the ISM 
simulations it is conceivable for coefficients $\alpha_{ij}$ 
to be anti-symmetric with respect to the galactic mid-plane, 
owing to the mirror asymmetry of vertically stratified helical 
turbulence. Similarly for $\eta_{ij}$ a symmetric vertical 
profile can be reasonably expected due to statistical symmetry 
of turbulent kinetic energy on either sides of the mid-plane.
Such symmetries are explicitly seen to obtain in closure 
calculations of the dynamo coefficients \citep{kr_rad_book,
RKR03,BS05,anv_kan_book}. Hence, a prior on dynamo coefficients 
can be formally incorporated in the aforementioned algorithm of 
IROS described in \sref{sec:iros_method}, and we apply it to the 
same data of ISM simulations as follows. We will refer to this 
method as 'constrained IROS' or `cIROS'.

Firstly, we set the zeroth level dynamo coefficients $\alpha_{
ij}(z)$ and $\eta_{ij}(z)$ to zero as before. Then {\it Step 1}, 
(\sref{sec:step1}) of the $k^\mathrm{th}$ refinement of IROS is 
modified as follows in cIROS: The vertical profiles of (the $k^
\mathrm{th}$ order) estimates of dynamo coefficients $^k\alpha_{
ij}( z ) $ and $^k\eta_{ij}( z)$, (its full functional form) is 
first determined by again keeping each of these {\it functions} 
non zero, turn by turn, and using \eref{eq:emf_local} at each 
time $t_i$. The functional form of the coefficients $^k\alpha_{
ij}(z)$ are expanded in terms of Legendre polynomials of odd 
degrees while the $^k\eta_{i j}$'s are expanded using ones with 
even degrees, and the expansion coefficients are used to obtain 
the fits 
\begin{align}
{f}_k^{\alpha_{ij}}\left(z\right)& = \sum_{\ell=0}^{n} a_{2\ell+1}\, P_{2\ell+1}\left(z/L_z\right),
\end{align}
and
\begin{align}
{f}_k^{\eta_{ij}}\left(z\right) &= \sum_{\ell=0}^{n} a_{2\ell}\, P_{2\ell}\left(z/L_z\right).  
\end{align} 
Here, $a_m$ are the fitting parameters\footnote{These parameters 
are determined by using the orthogonality property of Legendre 
polynomials}, $P_{m}\left(z/L_z\right)$ the Legendre polynomials 
of degree $m$, and $L_z$, the vertical extent of the simulation 
domain in $z$ direction ($\sim 2$\kpc). Thus ${f}_k^{\alpha_{ij}
}\left(z\right)$ is an odd function of $z$, while ${f}_k^{\eta_{
ij}}\left(z\right)$ is even in $z$ as required by our prior. The 
choice of $n$, the number of polynomials to be included in the 
fit, is a free parameter. Then to determine the best fitted dynamo 
coefficient, the chi-square errors associated with ${f}_k^{\alpha_{ij}}
\left(z\right)$ and ${f}_k^{\eta_{ij}}\left(z\right)$, 
denoted as $\chi^2_{il}$, are compared. Chi-square errors here are 
then defined as
\begin{align}
    \chi^2_{i0} &= \sum_z \sum_t \left(\mean{\emf}_i(z,t) - {f}_k^{\alpha_{ix}}(z)\,\mean{B}_x(z,t) \right)^2,\nonumber\\
    \chi^2_{i1} &= \sum_z \sum_t \left(\mean{\emf}_i(z,t) - {f}_k^{\alpha_{iy}}(z)\,\mean{B}_y(z,t) \right)^2,\nonumber\\
    \chi^2_{i2} &= \sum_z \sum_t \left(\mean{\emf}_i(z,t) + {f}_k^{\eta_{ix}}(z)\,\mean{J}_x(z,t) \right)^2,\nonumber\\
    \chi^2_{i3} &= \sum_z \sum_t \left(\mean{\emf}_i(z,t) + {f}_k^{\eta_{iy}}(z)\,\mean{J}_y(z,t) \right)^2.
\label{eq:chi_sq_iros_new}
\end{align}
Note here, that these errors are not only summed over time but also 
summed over $z$ and thus relate to the overall shape of the vertical 
profiles of dynamo coefficients. The determination of the best fit 
coefficient in cIROS is not therefore made separately at each $z'$, 
as opposed to that in previous case give in \eref{eq:chi_sq_iros_old}.

Further, in {\it Step 2} (\sref{sec:step2}) the contribution to the 
EMF associated with the least chi-squared weighted with the loop-gain 
factor (that is, either $\epsilon\,{f}_k^{\alpha_{ij}}\mean{B}_j$ or 
$-\epsilon\,{f}_k^{\eta_{ij}}\mean{J}_j$) is subtracted from $\mean{
\emf}_i$. {\it Step 3}, {\it step 4} and {\it 5} are then performed 
the same way as described in \sref{sec:step3}, \sref{sec:step4} and 
\sref{sec:step5}. Fitting at each $z$ is now not necessary as the 
$z$ dependence of the coefficients is already taken into account in 
the polynomial fit. The resulting vertical profiles of the dynamo 
coefficient are
\begin{align}
\alpha_{ij}(z) &= \epsilon\,\,\sum_{r=0}^{R}\,{f}_r^{\alpha_{ij}}(z),\\     
\eta_{ij}(z) &= \epsilon\,\,\sum_{r=0}^{R}\,{f}_r^{\eta_{ij}}(z).
\end{align}
We discuss the results of this analysis and compare the performance 
of cIROS on the determination of dynamo coefficients in 
\sref{sec:results_cIROS}. Additionally we test these two methods 
against the mock data of a noisy EMF in the following 
\sref{sec:mock_data}.

\section{Testing the IROS method using mock data}
\label{sec:mock_data}
We test the ability of the regular and the constrained IROS method 
to correctly estimate the profiles of dynamo coefficients by 
treating with it the mock data of the EMF generated from the mean 
magnetic fields. For this purpose we first prescribe the vertical 
profiles of mock dynamo coefficients $\alpha_{ij}^m(z)$ and $
\eta_{ij}^m(z)$, which are shown with black dot-dashed lines in
\fref{fig:mock_data}, and use the time evolution data of the 
vertical profiles of mean fields and currents ($\mean{\BB}(z,t)$ 
and $\mean{\JJ}(z,t)$) from the ISM simulations (discussed in 
\sref{sec:simulations}) to construct also the $z-t$ profiles of 
EMF using \eref{eq:emf_local}. Furthermore to construct mock data 
of the noisy EMF $\mean{\emf}^m(z,t)$, a uniform random 
distribution of noise of increasing power is added on the top of
it, such that the local ratios of noise to signal range from $10
\%$  to $300\%$. We note here also that the contribution of $
\alpha_{ij}^m(z)$ to the $\mean{\emf}^m(z,t)$ is taken to be much 
larger compared to that from  $\eta_{ij}^m(z)$. This is to ensure 
the efficacy of IROS to still determine the rest of the coefficients 
in situation where only few coefficients contribute overwhelmingly 
to the EMF. (Thus our mock data test is much more stringent than 
that required by the actual data).

\begin{figure}
    \centering
    \includegraphics[width=0.9\linewidth]{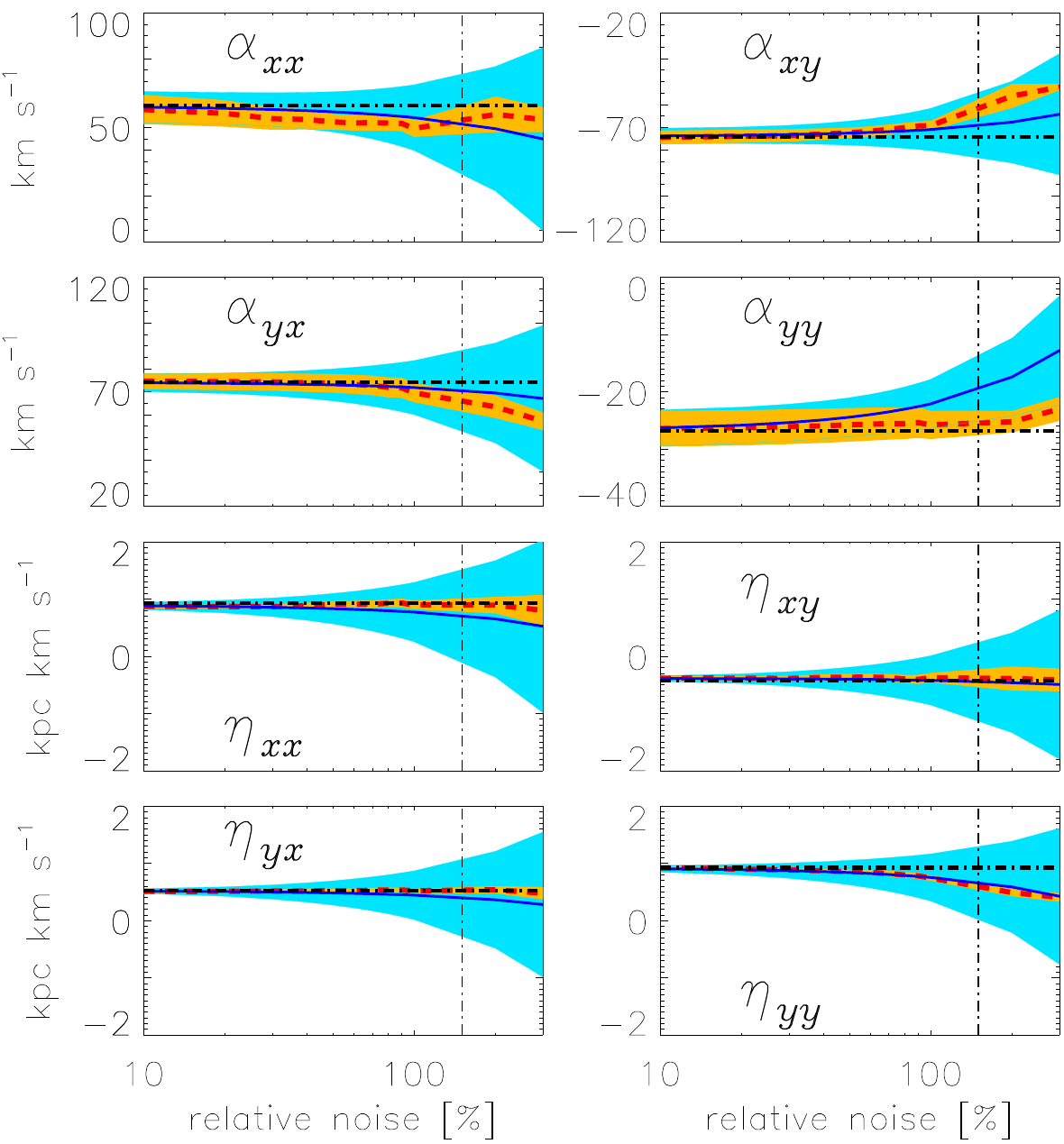}
    \caption{Plotted in dark blue-solid line style are the 
        estimates of the dynamo coefficients, averaged 
        between $z=-1$\kpc to -0.7 \kpc, as functions of 
        imposed error on the EMF, calculated with the IROS 
        method. The associated variance shown with light 
        blue colour. With the red dashed line we plot the 
        dynamo coefficients averaged over the same $z$ 
        domain, but estimated using the cIROS method along 
        with the corresponding variance shown in orange. 
        The black dot-dashed line are the imposed dynamo
        coefficients that make the mock EMF data. The 
        vertical dot-dashed line corresponds to the $150
        \%$ 
        of added noise to the mock EMF, to which
        \fref{fig:mock_data} corresponds.}
    \label{fig:noise_lev_comparison}
\end{figure}

We find that the IROS method in its unmodified version can determine 
most dynamo coefficients with accuracy up to the noise level of $\sim 
50\%$, and looses its efficacy to determine especially the $\eta_{ij}
$ beyond the noise level of $\sim50\%$, whereas with the constrained 
IROS or cIROS the coefficients can be determined reasonably up to a 
noise level of $\sim 200\%$. This is demonstrated in 
\fref{fig:noise_lev_comparison} where we can see that errors (shown 
in colour light blue) in the determinations of $\eta_{ij}$ with the 
regular IROS are overwhelmingly large beyond the noise level of $
\sim50\%$. While with cIROS method they are well constrained, as 
shown with the colour orange.

We want to clarify here the reason for using the mean-field data 
from the direct MHD simulations to construct the mock data of EMF, 
rather than one realised in the one dimensional dynamo simulations. 
The components of mean-field produced in pure $\alpha-\Omega$ dynamo 
simulations are completely correlated, and it is the difficult to
determine the associated dynamo coefficients separately, with any 
least-square minimisation method including IROS. Another source of 
error in determining dynamo coefficients from the data of direct 
simulations is additive noise in the EMF, which can however be 
handled with these methods.

\begin{figure}
    \centering
    \includegraphics[width=0.95\linewidth]{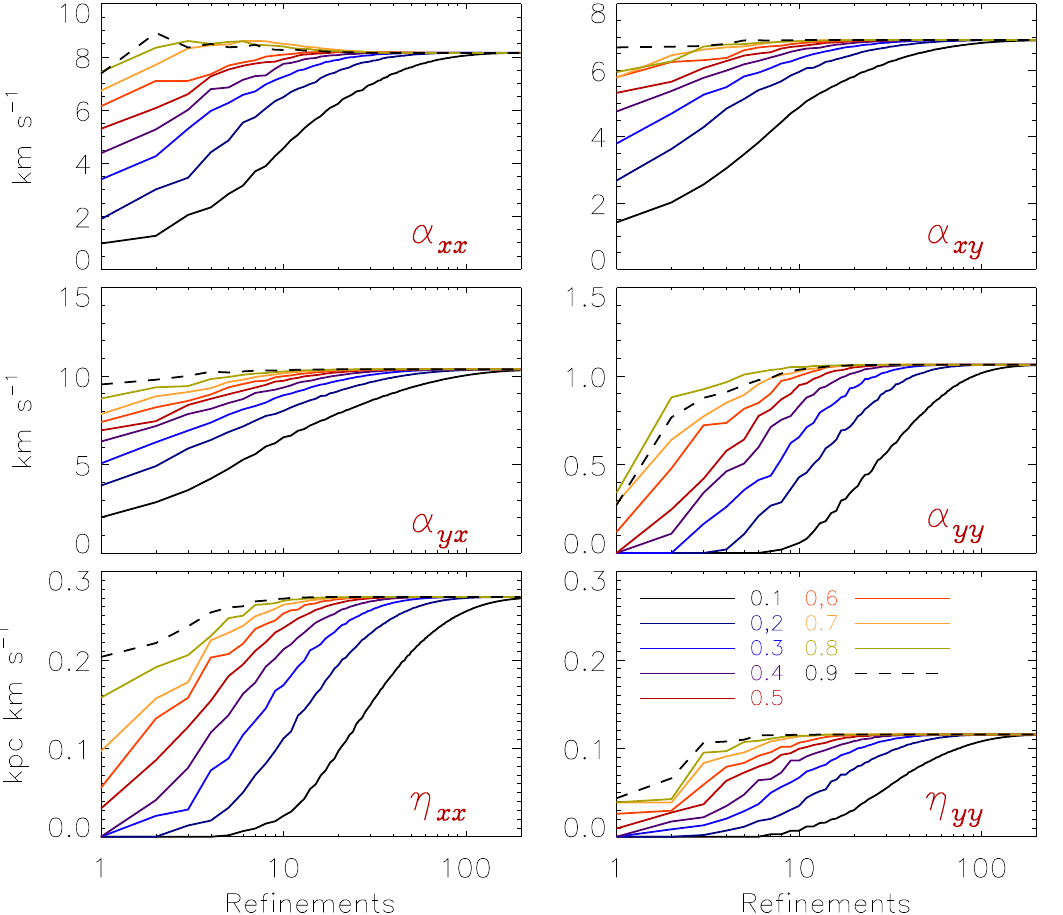}
    \caption{The obtained values of the dynamo coefficients as a function 
            of the number of refinements. Shown in different colours are 
            different loop gains $\epsilon$ ($10\%$ to $90\%$); see the 
            colour code in the lower right panel. Dynamo coefficients are
            calculated at a specific height ($\sim$700\pc), using the 
            local version of IROS.}
    \label{fig:coeff_converge_iros_loc}
\end{figure}

\section{results}
\label{sec:results}

\subsection{IROS inversion using the local EMF representation}
\label{sec:results_IROS_loc}
We apply the method discussed in \sref{sec:iros_method} to the data 
of MHD simulations of ISM described in \sref{sec:simulations}, and 
obtain the associated dynamo coefficients. We first extract the time 
series of the EMF, mean-fields and mean currents from the simulations 
and split them in nine different sections by successively skipping 
eight points. For each of these realisations we obtain all dynamo 
coefficients, using the aforementioned IROS method. In right hand 
panel of \fref{fig:cofficients_local} with dark blue solid lines we 
have shown the average of these nine outcomes, along with associated 
errors using light blue shades, which are obtained from the variance 
in these nine results. Here we have used the loop gain $\epsilon$ of 
$0.3$ and a hundred levels of refinements in total ($R=100$). 
Additionally these simulations have already been analysed using 
local SVD method. In the left hand panels of the same figure we plot 
the estimates of dynamo coefficients obtained using SVD with black 
dotted lines, along with their respective errors obtained in the 
same way by a green shade. It can be seen that both these methods
give very similar results although employing very different 
algorithms in detail.

We also perform the same exercise for different values of $\epsilon$
and track the values of refined dynamo coefficients as a function 
of refinements. It appears that as we refine these coefficients in 
IROS, they converge to their true values logarithmically. The rate 
of convergence is unsurprisingly faster for the larger values of 
loop gain $\epsilon$. We demonstrate this in 
\fref{fig:coeff_converge_iros_loc}, for the local version of IROS 
method, where we plot the magnitudes of various dynamo coefficients 
at $\sim800$\pc, as functions of refinements, with different line 
styles indicating the various loop gains as indicated in the 
lower-right hand panel. 

\subsection{cIROS inversions using prior constraints}
\label{sec:results_cIROS}
\begin{figure}
    \centering
    \includegraphics[width=\linewidth]{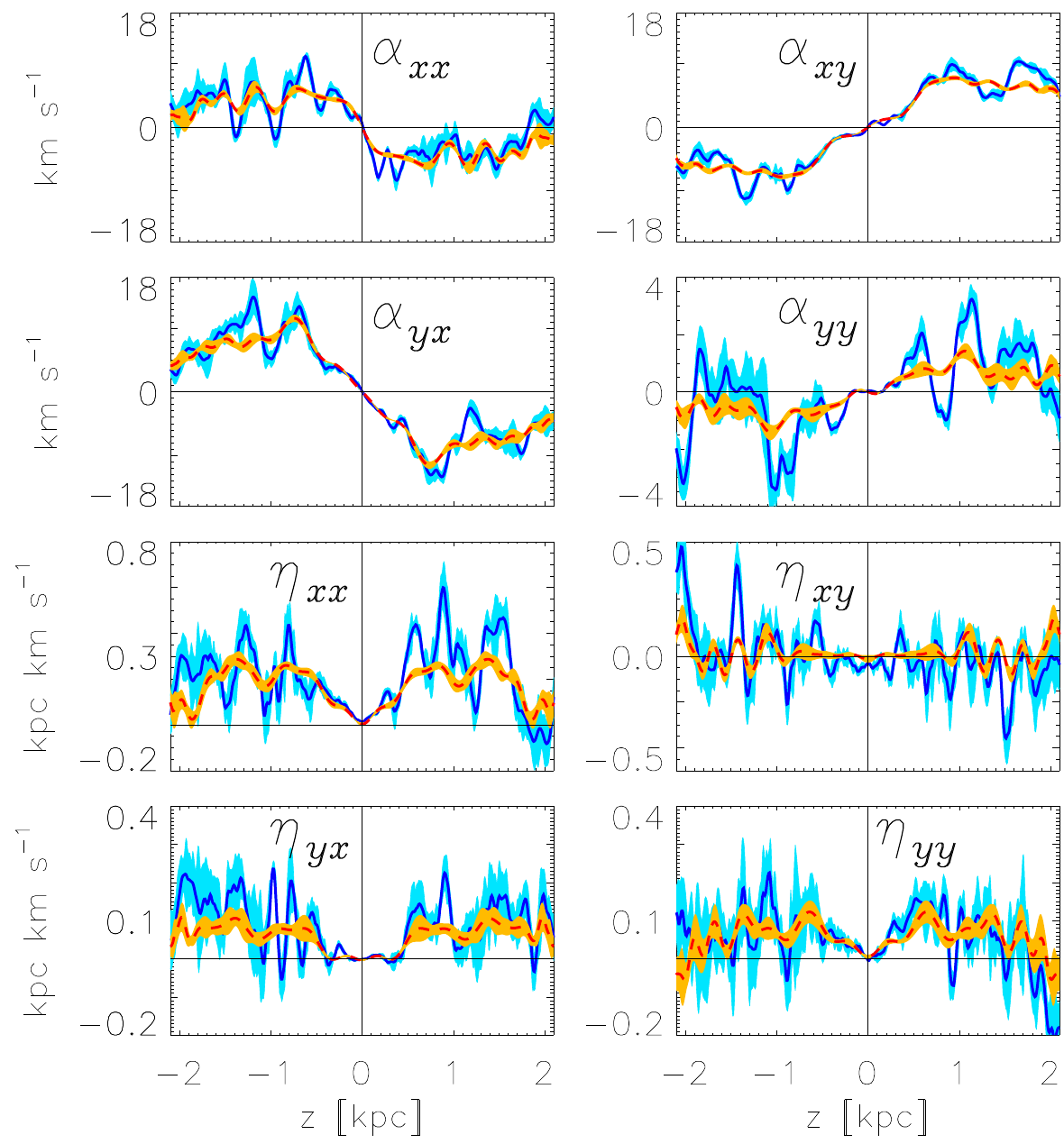}
    \caption{Shown with red dashed lines are the vertical 
        profiles of dynamo coefficients computed using the 
        cIROS method. Associated errors are shown with 
        orange shaded regions. While the dark blue solid 
        lines show the dynamo coefficient profiles computed 
        with the IROS method (same as in the left hand 
        panels of \fref{fig:cofficients_local}), along with 
        their associated errors are shown with light blue 
        shaded regions.}
    \label{fig:old_vs_new_iros}
\end{figure}

In a similar way we use the constrained IROS method (\sref{sec:cIROS}) 
to invert the local $\mean{\emf}$ representation. We use the same ISM 
simulations data discussed in \sref{sec:simulations} and using $n=40$ 
even and odd Legendre polynomials to fit the dynamo coefficients during 
each cIROS iteration, and obtain their vertical profiles. These are 
shown in \fref{fig:old_vs_new_iros}. With red dashed lines we show the
profiles obtained with the constrained IROS method along their errors 
shown in the orange shaded region. While with the blue solid lines we 
show the same with the unconstrained IROS method described in 
\sref{sec:iros_method}. We see that these two methods of inversions 
give qualitatively similar results, although the $\alpha_{ij}$ and 
$\eta_{ij}$ coefficients determined using cIROS after imposing priors 
are firstly antisymmetric and symmetric respectively, with respect to 
the midplane, and secondly their profiles are smoother with smaller 
errors. 

Even without applying these constraints, the rough tendencies of even 
or odd parities of the vertical profiles of the most coefficients can 
already be seen in the results of IROS (and SVD) methods 
\fref{fig:coeff_converge_iros_loc}, which is to be expected for the 
SN driven ISM turbulence. We note here, however, that the priors used 
for $\alpha_{ij}$ and $\eta_{ij}$ are specific to the ISM simulations 
where the parities of the coefficients can be inferred from underlying
physics. These need not be valid for other astrophysical systems. 
However any other physically informed constraint specific to a 
particular system or more general priors can be similarly incorporated 
in the framework of IROS.

\section{Conclusions}
\label{sec:conclusion}
Estimation of the dynamo coefficients associated with MHD simulations 
is crucial in understanding the large-scale dynamo effect as well as 
in quantifying the connection of MHD turbulence to the evolution of 
large-scale magnetic fields. This is usually accomplished using either 
the test-field method, which requires multiple additional induction
equations to be solved along with the DNS, or using the post-processing
methods based on regression which tend to be relatively faster. The 
SVD method for example directly solves \eref{eq:emf_local} in 
post-processing, by assuming the constancy of dynamo coefficients in 
time and inverting \eref{eq:emf_overdetermined_local} instead. Problems 
associated with directly inverting \eref{eq:emf_local}, usually stem 
from having a statistical correlation between $\mean{\BB}$ and $\mean{
\JJ}$ (for example in $\alpha^2$-dynamos, $\mean{\JJ}\sim\nabla\times
\mean{\BB} \sim k\,\mean{\BB}$, which further introduces correlations 
in $\alpha_{ij}$ and $ \eta_{ij}$), or having one (or more) of the 
components of $\mean{\BB}$ or $\mean{\JJ}$ contribute overwhelmingly 
to the EMF. The latter issue in particular leads to the underestimation 
the dynamo coefficients associated with the components of the mean field 
or current that contribute negligibly to the EMF or are comparatively 
noisier.

In this paper we present a new post processing tool, IROS, which mitigates 
this later problem, apart from having a number of other advantages. In 
IROS the components of EMF are successively fit against the individual mean 
field and current components, and the estimates of dynamo coefficients 
obtained in the fits are refined iteratively by removing the contribution 
of the best fitted parameters from the EMF, thus circumventing the 
aforementioned issue. Following a similar algorithm we further extended 
this method to also invert the non-local representation of EMF to determine 
the components of the non-local kernel $\mathcal{K}_{ij}$, as described in 
\aref{sec:nonlocal}. Significantly, we also show that it is possible in 
IROS to impose reasonable prior constraints on the dynamo coefficients. 
This constrained IROS method (``cIROS''), makes an appropriate modification 
to the criterion of the best fit at each iteration of the IROS refinement. 
As an example, we impose the priors on the vertical profiles of dynamo 
coefficients, such that $\alpha_{ij}(z)$ and $\eta_{ij}(z)$ with respect 
to the mid-plane ($z=0$) are antisymmetric and symmetric respectively. 
This was done simply by expanding the dynamo coefficient profiles in terms 
of either odd or even Legendre polynomials and modifying the definitions 
of chi-squared errors to measure the goodness of the EMF fit with respect 
to these Legendre expansions, as described in \sref{sec:cIROS}. We note 
here that a different set of prior constraints specific to the system can 
be incorporated in the same way with appropriate definitions of chi-squared 
errors. Moreover a probabilistic framework can also be adopted to impose 
the said priors, self consistently. 

To have a reasonable validation of the IROS method we applied it to the
data of ISM simulations, which has been analysed before using the 
test-field method as well as the local and non-local variants of the 
SVD method. The vertical profiles of the dynamo coefficients recovered 
from the IROS and the SVD method are found to be largely consistent 
with each other (as shown in \fref{fig:cofficients_local} and 
\fref{fig:coefficients_non_local}). We also applied this method to the 
synthetic data with predefined vertical profiles of dynamo coefficients 
chosen such that only a few of the coefficients contributed largely to 
the EMF, and with varying levels of additive noise. We demonstrated that 
even with the noise level as high as 200\% of the EMF, using IROS we 
were still able to recover the dynamo coefficients with a reasonable 
accuracy, and considerably better than the SVD method. We have applied 
these methods  to compute the dynamo coefficients associated with other 
simulations, with a reasonable degree of success. For example in 
\citep{prasun} we apply these methods for the local simulations of 
accretion disc, where the turbulence is driven by the magnetorotational 
instability. While in \citep{yasin} we use them in the simulations 
where the turbulence is driven by the magnetic buoyancy instability.

With this analysis we have shown that the IROS method could serve as a 
viable method to determine the dynamo coefficients. As a post-processing 
tool, it firstly has an advantage of being extremely computationally 
efficient compared to the test-field method, while also being more robust 
in handling the additive noise than the SVD. A conceivable disadvantage in 
the standard IROS (and also in the SVD), however, is that the determined 
dynamo coefficients get correlated when the components of mean-field and 
current are correlated as well. This is avoided in the test-fields method 
since the additional linearly independent test magnetic fields are also 
evolved along with the MHD simulations. In this respect, it is useful to 
have a possibility, as in the cIROS, of imposing prior constraints on 
dynamo coefficients to break the degeneracy between $\alpha_{ij}$ and 
$\eta_{ij}$. It would also be useful to extend the the cIROS to invert 
the integral or kernel representation of EMF and to impose prior 
constraints on the coefficients of the non-local kernel. 

\section*{Acknowledgements}
We thank Dipankar Bhattacharya for valuable insights and for a detailed tutorial 
on IROS. We also thank Neeraj Gupta, Maarit J. Korpi-Lagg, and Matthias Rheinhardt 
for their valuable insights and explanations. J.S.~and A.B.~acknowledge the support 
by the Swiss National Science Foundation under Grant No.\ 185863.

\section*{Data Availability}
The data underlying this article will be shared on 
reasonable request to the corresponding author.


\bibliographystyle{mnras}
\bibliography{main}



\appendix
\section{Non-local IROS method}
\label{sec:nonlocal}
The turbulent EMF at any particular $z = z'$ can equivalently 
be expressed as a linear function of mean field components in 
its local $\zeta$ neighbourhood, through a non-local kernel $
\mathcal{K}_{ij}$ as follows.
\begin{align}
\mean{\emf}_i(z')=\int_{-\infty}^{\infty} \mathcal{K}_{ij} \left(z,\zeta\right) \mean{B}_j\left(z-\zeta\right) {\rm d}\zeta.
\label{eq:emf_non_local}
\end{align}
This representation of EMF allows one to express the regular
dynamo coefficients $\alpha_{ij}$, $\eta_{ij}$ as the zeroth 
and first moments of $K_{ij}$ respectively. Additionally, 
from the numerical perspective it has an advantage of not 
being dependent directly on the coefficients of current, 
avoiding any errors in fitting that may enter in the 
computation by having to numerically differentiate $\mean{
\BB}$. Furthermore, it also accounts for the non-local 
contribution of mean-fields to the turbulent EMF. In 
\citep{non_loc_svd}, by varying the width of the local
window, we have established that the non-local 
contributions up to $\sim 50$\pc contribute significantly 
to the EMF. 

To express \eref{eq:emf_non_local} in a form that can be 
solved by least square minimisation, we represent it first 
in a summation form,
\begin{align}
    \mean{\emf}_i = \sum_{l=-m}^{m}\, \mathcal{K}_{ij}(z',\zeta_l)\,\mean{B}_j(z'-\zeta_l)\,\delta,
    \label{eq:emf_non_local_sum}
\end{align}
where the $\delta$ in the size of grid, and the limit $l$ 
represent the number of points we want to consider in the 
local neighbourhood of $z'$. $\mathcal{K}_{ij}$, typically 
vanish as $ \zeta_l $ approaches the correlation scale of 
turbulence, which can be determined empirically by 
considering various values of $l$. In this particular 
example $l=6$ spans the local neighbourhood of $z'$ as $6
\delta(\sim 50\pc)$ approximately corresponds to the 
scale-length of SN driving. 

We rewrite \eref{eq:emf_non_local_sum} then in a matrix 
form (similar to \eref{eq:emf_overdetermined_local}) as 
follows, 
\begin{align}
        \mathbf{y}_i \left(z',t\right)= \mathbf{A}\left(z'\right)\, \mathbf{x}_i \left(z',\zeta_l\right),
    \label{eq:emf_overdetermined_non_local}
\end{align}
where, $\mathbf{A}$ is a $N\times 2(2l+1)$ matrix, comprised of 
$\mean{B}_x$ and $\mean{B}_y$ in the local neighbourhood of $ z 
= z'  $, with $\zeta_l$ being the neighbourhood variable, and $
\mathbf{ x}_i(z', \zeta) $ the kernel coefficients, while the $
\mathbf{y}_i(z',t)$ is the column vector with EMF (component $i
$) time series at $ z=z'$. 
\begin{figure}
    \centering
    \includegraphics[width=0.9\linewidth]{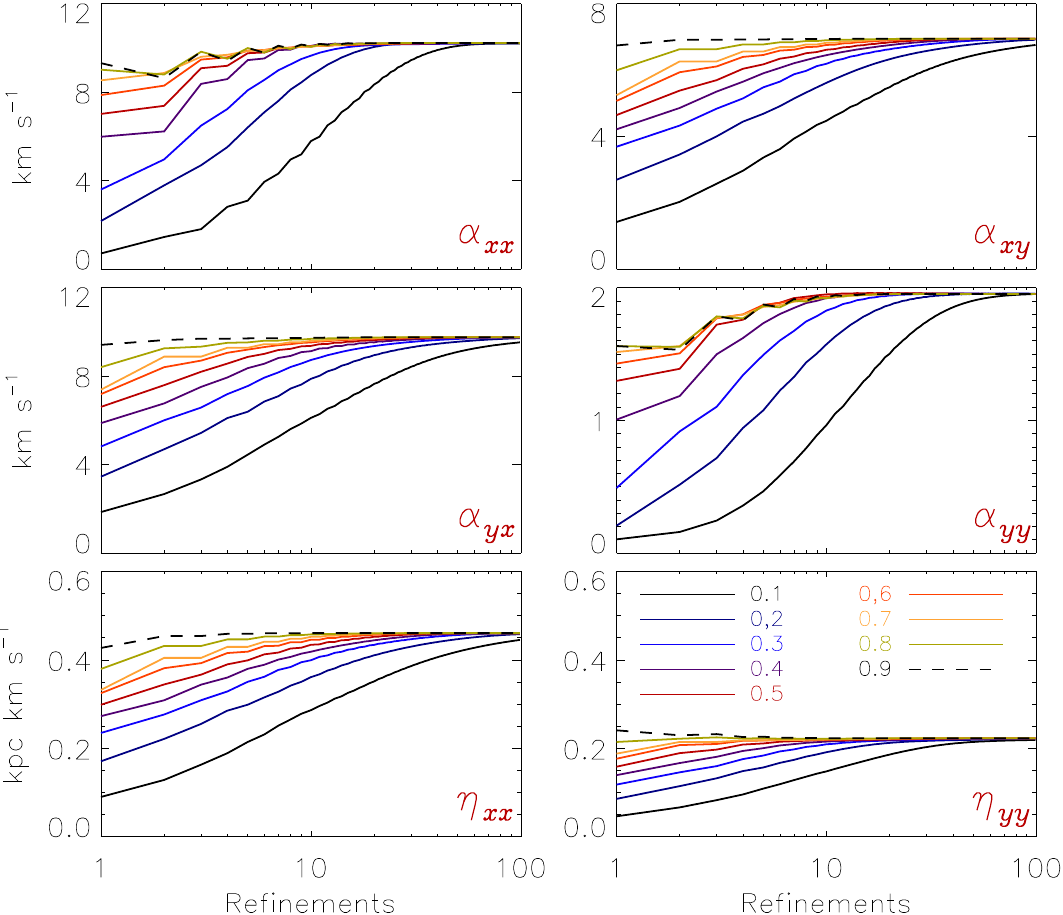}
    \caption{Same as \fref{fig:coeff_converge_iros_loc} but 
    for the dynamo coefficients computed with the non-local 
    version of IROS.}
    \label{fig:coeff_converge_iros_non_loc}
\end{figure}
Specifically,
\begin{align}
    \mathbf{y}_i(z',t) &=
        \begin{bmatrix}
            \mean{\emf}_i\left(z',t_1\right)\\
            \mean{\emf}_i\left(z',t_2\right)\\
            \vdots\\
            \mean{\emf}_i\left(z',t_N\right)
        \end{bmatrix},\\\nonumber\\
    \mathbf{x}_i(z',\zeta_l) &=        
        \begin{bmatrix}
            \mathcal{K}_{ix}\left(z'-l\delta\right)\\
            \vdots\\
            \mathcal{K}_{ix}\left(z'+l\delta\right)\\
            \\
            \mathcal{K}_{iy}\left(z'-l\delta\right)\\
            \vdots\\
            \mathcal{K}_{iy}\left(z'+l\delta\right)
        \end{bmatrix},\\\nonumber\\
    \mathbf{A}^{\intercal}\left(z'\right) &= 
        \begin{bmatrix}
            \mean{B}_{x}\left(z'-l\delta,t_1\right) 
            &\mean{B}_{x}\left(z'-l\delta,t_2\right) 
            \hdots\mean{B}_{x}\left(z'-l\delta,t_N\right)\\
            \vdots                                
            &\vdots                                     \\
            \mean{B}_{x}\left(z'+l\delta,t_1\right) 
            &\mean{B}_{x}\left(z'+l\delta,t_2\right) 
            \hdots\mean{B}_{x}\left(z'+l\delta,t_N\right)\\
            &                                         \\           
            \mean{B}_{y}\left(z'-l\delta,t_1\right) 
            &\mean{B}_{y}\left(z'-l\delta,t_2\right) 
            \hdots\mean{B}_{y}\left(z'-l\delta,t_N\right)\\
            \vdots                                
            &\vdots                                     \\
            \mean{B}_{y}\left(z'+l\delta,t_1\right) 
            &\mean{B}_{y}\left(z'+l\delta,t_2\right) 
            \hdots\mean{B}_{y}\left(z'+l\delta,t_N\right)
            \end{bmatrix}  
\end{align}
$\mathcal{K}_{ij}$ components in this formulation relate to the 
regular dynamo coefficients $\alpha_{ij}$ and $\eta_{ij}$ through 
following relations \citep[see e.g.][]{non_loc_svd},
\begin{align}
    \alpha_{ij}\left(z\right) &= \int_{\zeta_l}\,\mathcal{K}_{ij}\left(z,\zeta_l\right)\,d\zeta_l \nonumber\\\nonumber\\
    \begin{bmatrix}
    \eta_{xx}   &   \eta_{xy}\\
    \eta_{yx}   &   \eta_{yy}
    \end{bmatrix}\left(z\right)&=
    \int_{\zeta_l}
    \begin{bmatrix}
    -\mathcal{K}_{xy}\left(z,\zeta_l\right)  &\mathcal{K}_{xx}\left(z,\zeta_l\right)\\
    -\mathcal{K}_{yy}\left(z,\zeta_l\right)  &\mathcal{K}_{yx}\left(z,\zeta_l\right)
    \end{bmatrix}\,\zeta_l\,d\zeta_l
\end{align}
To obtain then the $\mathbf{x}_i$ such that it minimises the residual  
$\mathcal{R}=(\mathbf{y}_i-\mathbf{A}\mathbf{x}_i)^2$, we follow the 
same IROS algorithm described in \sref{sec:iros_method}. With a 
slight modification described as follows. In {\it Step 1} we set all 
the coefficients of kernel $\mathcal{K}_{ij}$ to zero. Then, at any 
particular $z=z'$, we fit the time series $\emf_i(z',t)$ separately 
against $ \mean{B}_x(z'-\zeta)$ and $ \mean{B}_y (z'-\zeta)$ in the 
entire local $\zeta$ neighbourhood. That is, we fit $\mathbf{y}_i(z'
,t)$ against $\mathbf{A}^x(z')$ and $\mathbf{A}^y(z') $ separately, 
which contain the first and later $(2l+1)$ columns of $\mathbf{A}$,
\begin{align}
    \mathbf{A}^i(z') = 
    \begin{bmatrix}
    \mean{B}_{i}\left(z'-l\delta,t_1\right) & \mean{B}_{i}\left(z'+l\delta,t_1\right)   \\
    \vdots                                  & \vdots                                    \\
    \mean{B}_{i}\left(z'-l\delta,t_N\right) & \mean{B}_{i}\left(z'+l\delta,t_N\right)
    \end{bmatrix},
\end{align}
and obtain $^0\mathcal{K}_{ix}(z',\zeta)$ and $^0\mathcal{K}_{iy}
(z',\zeta)$, the zeroth level estimates of kernel coefficients .
Consequently the the definition of the chi-square errors is 
modified as
\begin{align}
    \chi_{i0}^2(z') = \sum_t \left(\emf_i(z') - \sum_{j=-m}^m {^0\mathcal{K}_{ix}}(z',j\delta)\,\mean{B}_x(z'-j\delta) \right)^2,\\
    \chi_{i1}^2(z') = \sum_t \left(\emf_i(z') - \sum_{j=-m}^m {^0\mathcal{K}_{iy}}(z',j\delta)\,\mean{B}_y(z'-j\delta) \right)^2.
\end{align}
Further steps are carried out the same way described in 
\sref{sec:iros_method}, and final estimate of kernel are obtained 
iteratively. The results of this analysis will also be discussed 
in \sref{sec:results}.
\begin{figure}
    \centering
    \includegraphics[width=\linewidth]{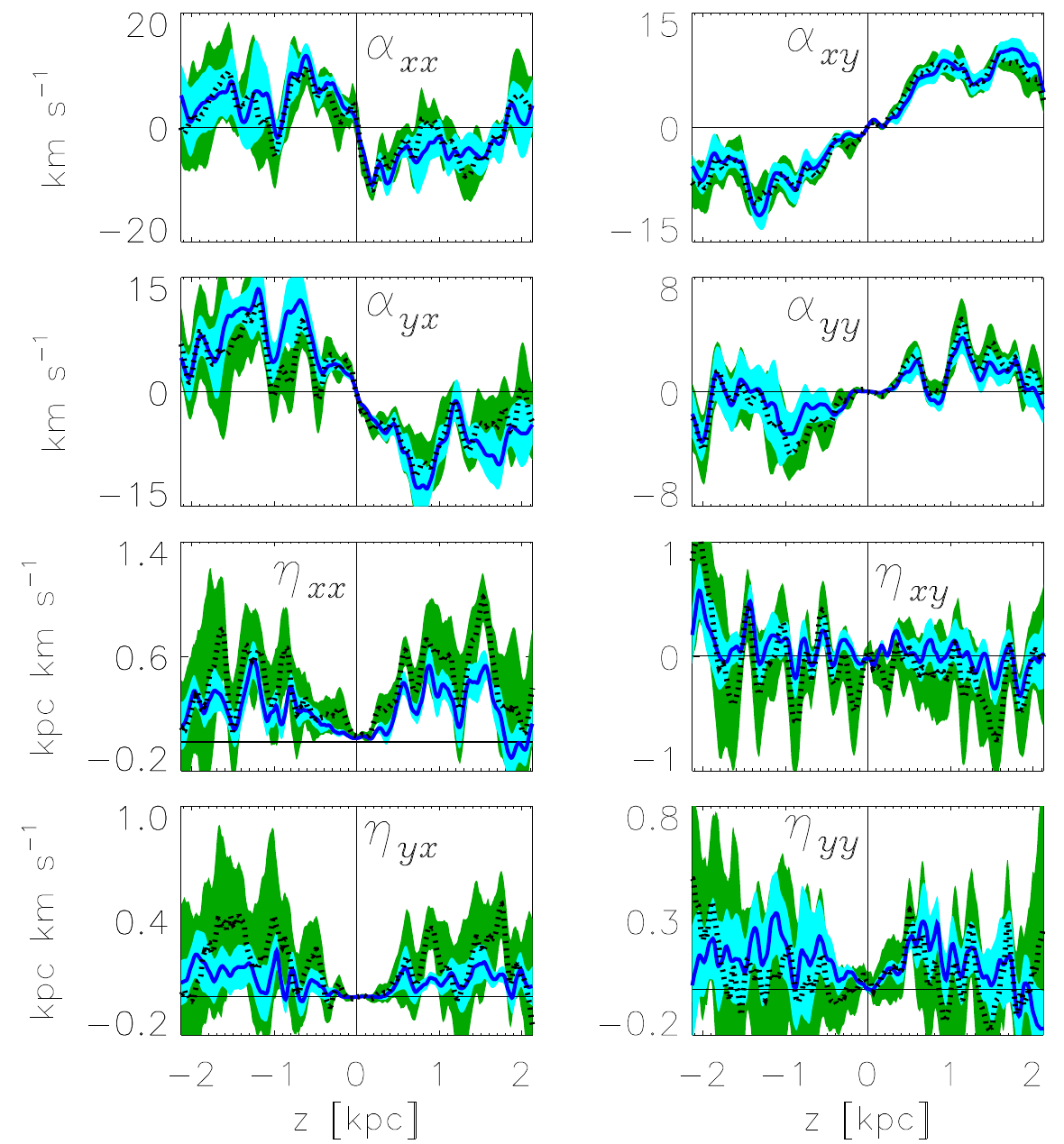}
    \caption{Same as \fref{fig:cofficients_local} except the 
        coefficients here are computed by applying IROS 
        algorithm to the non-local prescription of EMF. 
        With dark blue solid line we show the profiles of the 
        coefficients computed using IROS method with non-local
        EMF prescription. The associated $1-\sigma$ errors are
        shown with light blue regions. These coefficients are 
        also comparable with once calculated using non-local
        SVD method shown here with blick dotted lines and 
        associated errors in light green shades. Size of the 
        local neighbourhood considered here ranges from $\sim 
        -60$ to +60\pc about any particular $z$.}
    \label{fig:coefficients_non_local}
\end{figure}

\subsection{IROS inversions using the non-local EMF representation}
We now apply the method discussed in \sref{sec:nonlocal} to the same 
data. Vertical profiles of the zeroth and first moments of the dynamo 
coefficients (i.e. $\alpha_{ij}(z)$ and $ \eta_{ij}(z)$ respectively) 
are shown with dark blue solid lines in 
\fref{fig:coefficients_non_local}, along with $1-\sigma$ error 
intervals shown with light blue colour shades. Value of the loop gain 
$\epsilon$ and refinements used for the calculation are 0.3 and one 
hundred respectively. These profiles seem to compare well with the 
estimates of the same quantities made using non-local SVD method, 
shown with black dotted lines along with associated errors in green 
coloured shade. It can be seen by comparing the error bars in this 
figure, that the non-local IROS performs somewhat better than the 
non-local SVD method. The sensitivity of the results to a change in 
$\epsilon$ and the number of refinements is again tested and the 
results shown in \fref{fig:coeff_converge_iros_non_loc}.

\section{Flowchart of the IROS method}
\label{sec:flowchart}
\begin{figure}
    \centering
    \includegraphics[trim={4cm 4cm 4cm 0},clip,width=1.\linewidth]{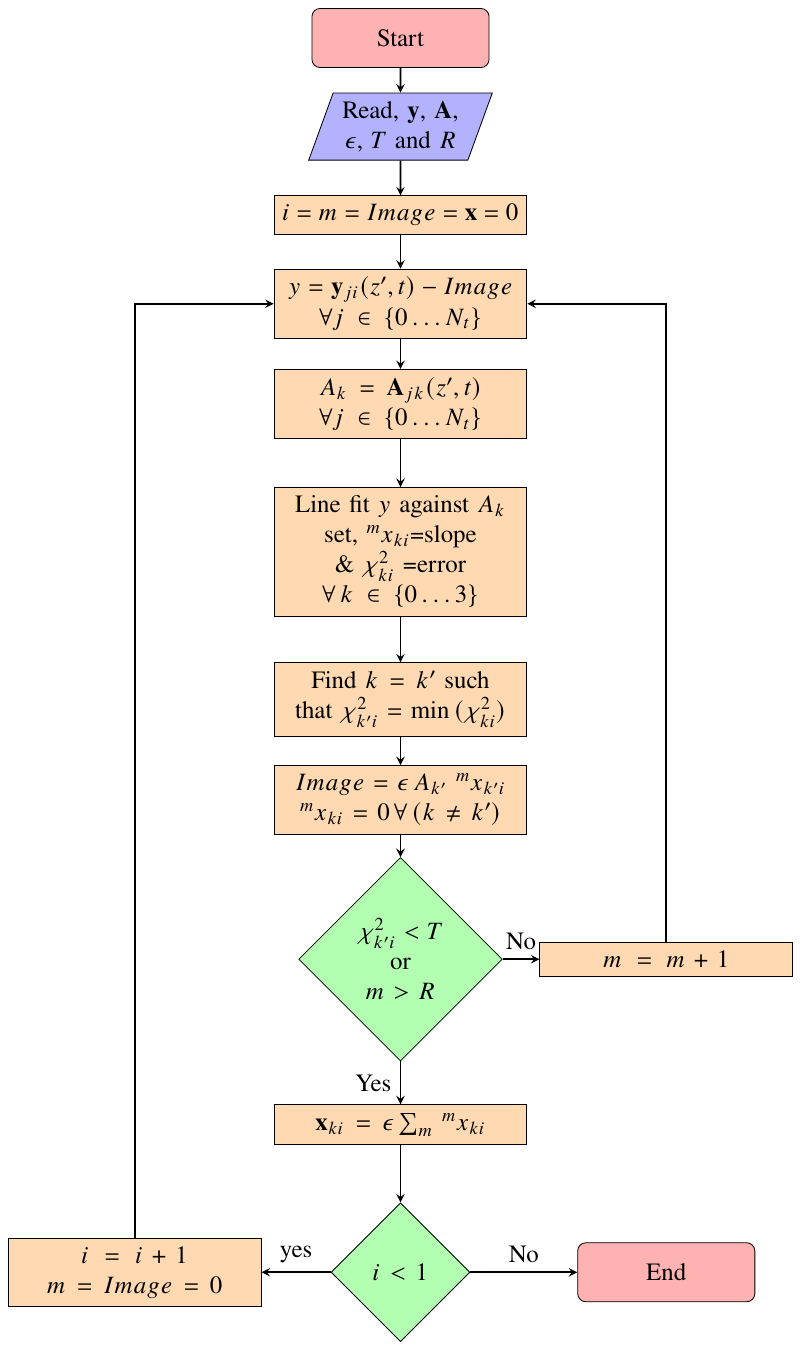}
    \caption{Flow chart representation of the local IROS algorithm.}
    \label{fig:flowchart}
\end{figure}

To further clarify the local IROS algorithm discussed in 
\sref{sec:iros_method} it can equivalently be represented as a
flowchart. For example, \fref{fig:flowchart} we show the flowchart 
of the local IROS method to determine the dynamo coefficients at $
z=z'$. To similarly represent the cIROS algorithm as a flowchart, a 
few modifications are made. Namely, instead of reading the data $y$ 
and $A_k$ at a particular $z=z'$ the entire $z$ profiles are read 
in, and after the line fitting procedure, the slopes $^mx_{ki}$ are 
expanded in terms of Legendre polynomials to calculate the 
chi-squared errors using \eref{eq:chi_sq_iros_new}. 


\bsp
\label{lastpage}
\end{document}